%% file: main.tex
\documentclass[12pt]{article}

\input{Preamble}

\date{}

\title{\textbf{Five-dimensional electrostatic black holes in a background field}}

\author{Fred Tomlinson\footnote{f.tomlinson@ed.ac.uk}
\\ \\ \small \sl School of Mathematics and Maxwell Institute for Mathematical Sciences, \\ \small \sl    University of Edinburgh, King's Buildings, Edinburgh, EH9 3FD, UK }

\begin{document}

\maketitle
\begin{abstract}

We consider $5$ dimensional electrostatic solutions to Einstein-Maxwell gravity with $2$ commuting spacelike Killing fields. Taking two distinct reductions from $5$ dimensions to a $3$ dimensional base space, we write the Einstein-Maxwell equations using some axially symmetric functions on $\mathbb{R}^3$. These equations can be viewed as arising from a harmonic map coupled to 3-dimensional gravity with the isometries of the target space of this map revealing a hidden $SL(2,\mathbb{R})$ symmetry of this sector of the theory. Depending on the choice of reduction this symmetry then gives rise to two different 1-parameter families of transformations corresponding to either charging a black hole or immersing it in a background electric field. We use these transformations to charge a static black Saturn and a static $L(n,1)$ black lens spacetime and by tuning the strength of the external field, we cure the conical singularities to give new regular solutions. Notably the electrified black lens generated is the first example of a regular black lens in Einstein-Maxwell gravity with topologically trivial asymptotics.
\end{abstract}
\newpage

\tableofcontents

\section{Introduction}

Equilibrium black holes in four-dimensional Einstein-Maxwell gravity are fully classified by their mass, charge and angular momentum. In particular the no-hair theorem states that any stationary axisymmetric and asymptotically flat (AF) black hole spacetime is isometric to a member of the Kerr-Newman family of solutions \cite{Chrusciel:2012jk}. If we consider gravity in higher dimensions, then no similarly general result is known. The strongest classification theorem is for static spacetimes which states that all black hole solutions must be members of the ($D$-dimensional) Reissner-Nordström family of solutions \cite{Gibbons:2002bh,Gibbons:2002ju,Kunduri:2017htl}. In this paper we will consider the more general class of stationary, multi-axisymmetric, AF solutions in $D$-dimensional Einstein-Maxwell gravity - a natural generalisation of the class of solutions governed by the no-hair theorem. It is notable that even with our somewhat restrictive assumptions, little is known about this theory. This is in stark contrast to the corresponding sectors of vacuum gravity and minimal supergravity which are governed by powerful uniqueness theorems \cite{Hollands:2007aj,Hollands:2008fm,Tomizawa:2009ua, Tomizawa:2009tb, Armas:2009dd, Hollands:2012cc, Armas:2014gga}. These theorems state that the solutions of the two theories are uniquely determined by their asymptotic charges together with so-called rod data - certain invariants of a solution defined on the horizons and axes of symmetry (see Section \ref{sec:background} for more details). These uniqueness theorems are proved using the fact that the two theories can be rewritten using a gravitating harmonic map to a coset space, however no such harmonic map description is known for Einstein-Maxwell gravity in more than four dimensions.

Along with these uniqueness theorems, another key consequence of this harmonic map description is the existence of hidden symmetries. Explictly, if the harmonic map is to a $G/H$ coset space, then the isometries of this target space are given by $G$ which in turn corresponds to symmetries of the original equations (see \cite{Clement:2008qx, Galtsov:2008zz} for reviews). The simplest example of these symmetries appears in stationary, axisymmetric, vacuum solutions in 4 dimensions where an Ernst system arises which has an $SL(2,\mathbb{R})$ symmetry \cite{Ehlers:1959aug}. To see this from the perspective of the harmonic map, one reduces to $3$ dimensions by quotienting out the orbits of the $U(1)$ symmetry corresponding to the axisymmetric Killing vector field (KVF). Similar results hold in various other theories of gravity: $D$-dimensional vacuum gravity has an $SL(D-2,\mathbb{R})$ symmetry \cite{Maison:1979kx}; $4$-dimensional Einstein-Maxwell gravity has an $SU(2,1)$ symmetry \cite{Kramer:1969wq}; $5$-dimensional minimal supergravity has a $G_{2(2)}$ symmetry \cite{Mizoguchi:1998wv}; $11$-dimensional supergravity has an $E_{8(8)}$ symmetry \cite{Julia:1980gr,Mizoguchi:1997si}. However, as mentioned above, it is unknown how to write the equations for Einstein-Maxwell gravity for $D>4$ in terms of a harmonic map to some coset space and so symmetries of the theory are more obscured. It is important to note that $5$-dimensional minimal supergravity reduces to Einstein-Maxwell gravity when the Chern-Simons term vanishes which in particular occurs when there is no magnetic field. Therefore purely electric Einstein-Maxwell gravity inherits the uniqueness theorem and some of the symmetries of the supergravity theory.

Although there are uniqueness results known in five-dimensional, AF vacuum gravity and minimal supergravity, the space of solutions is still not fully understood. For example it is known that black hole horizons in five dimensions must either have $S^3$, $S^2 \times S^1$ or $L(p,q)$ (lens space) topology \cite{Hollands:2007aj} however for the vacuum case only $S^3$ and $S^2 \times S^1$ horizon black holes are known. In fact it can be shown that the simplest vacuum lens space topology black hole solution compatible with the uniqueness theorem is always singular \cite{Lucietti:2020phh,Lucietti:2020ltw}, providing some evidence that vacuum black lenses do not exist in general. Similarly there are no known non-extreme AF black lens solutions in minimal supergravity though there are many extremal examples of black lenses which have been constructed \cite{Kunduri:2014iga,Tomizawa:2016kjh}, fitting into a general classification of supersymmetric black holes \cite{Breunholder:2017ubu}). Part of the motivation for studying Einstein-Maxwell gravity in this paper is as a toy model to understand solutions that haven't yet been written down in these other theories. To do this we will consider static solutions and then charge them using the hidden symmetries discussed above to give new electrostatic solutions which will preserve the rod structure of the original. These new charged solutions circumvent the static uniqueness theorem since they are no longer AF and are instead embedded in an external electric background. Whilst this breaks asymptotic flatness, it still preserves the asymptotic topology of the metric, i.e. the constant time slices are still topologically $S^3$ at infinity.

A large class of static vacuum solutions is given by Weyl solutions. These are five-dimensional solutions with $2$ orthogonal axial KVFs, that can be trivially constructed out of axially symmetric harmonic functions on $\mathbb{R}^3$ \cite{Emparan:2001wk}. Using the static uniqueness theorem we know that AF solutions in this class are singular for all but the flat or Schwarzschild case, so general Weyl solutions must somehow be modified to construct something regular. One way to do this is by adding rotation, which can be achieved using the inverse scattering method. This is a method based on integrability that takes a seed Weyl solution and uses it to generate a more general solution using a particular "soliton" ansatz \cite{Belinsky:1971nt,Belinsky:1979mh,Pomeransky:2005sj}. A different approach is to add charge to these solutions to balance them. Again one can do this using inverse scattering \cite{Figueras:2009mc}, however there are other methods developed to do these charging transformations relying explicitly on hidden symmetries of the theory. In $11$ dimensions one can charge solutions using the U-duality of supergravity (equivalently the $E_{8(8)}$ symmetry we discussed above), which is also inherited by supergravity theories in lower dimensions through dimensional reduction (see e.g. \cite{Elvang:2004xi, Kunduri:2004da, Lu:2008ze} for applications to minimal supergravity).

In this paper we will develop charging transformations for biaxisymmetric, electrostatic solutions in five dimensions by using some $SL(2,\mathbb{R})$ hidden symmetries. We apply this transformation to the case of the black Saturn and a simple black lens. Charged black Saturns in Einstein-Maxwell gravity have already been constructed, for example a singular static charged solution \cite{Chng:2008sr} and a regular rotating solution with a dipole charge \cite{Yazadjiev:2007cd}. Black lens solutions can be constructed in a trivial way by taking a Schwarzschild solution and quotienting by an appropriate discrete subgroup of $SO(4)$ - this however affects the topology of the solution at infinity. Other than these solutions no black lenses have previously been constructed in Einstein-Maxwell gravity. The new regular charged solution that we derive gives the first example of a solution which is topologically asymptotically flat.

This paper is organised as follows: In Section \ref{sec:background} we discuss two different reductions from the $D$-dimensional theory to a $3$-dimensional base space. This allows us to write the Einstein-Maxwell equations in a convenient way adapted to either reduction. In Section \ref{sec:transformations} we specialise to the case of electrostatic solutions in five dimensions. Each of the two choices of reductions from the previous section lead to equations with different apparent symmetries. Exploiting these symmetries, we derive two different 1-parameter families of charging transformations. These combine to give a $2$-parameter family of transformations which charge a solution and then immerse it in a background electric field. In Section \ref{sec:examples} we apply this combined transformation to the black Saturn and a $L(n,1)$ black lens solution and, by appropriately tuning the strength of the external electric field, cure the conical singularities. We end with a discussion of these results and possible extensions in Section \ref{sec:disc}. In Appendix (\ref{ap:Weyl}) we discuss the charging of a general Weyl solution and write down the balance conditions that the charged solutions must satisfy.

\section{Background}
\label{sec:background}
We begin by considering a $D$-dimensional spacetime $(M,g,F)$ in Einstein-Maxwell gravity with action
\begin{equation}
\label{EMaction}
    S = \int_M R \star 1 - 2 F \wedge \star F.
\end{equation}
In addition we assume that the spacetime is AF, and possesses a stationary KVF $k$ and $D-3$ compatible axial KVFs $m_i$ ($i= 1, \dots , D-3)$, a result of which is that the isometry group has an $G:=\mathbb{R}\times U(1)^{D-3}$ subgroup. These symmetries allow us to write the metric in Weyl-Papapetrou coordinates \cite{Emparan:2001wk, Harmark:2004rm,Ida:2003wv}
\begin{equation}
    g = g_{AB}(\rho,z) \td x^A \td x^B + e^{2\nu(\rho,z)}(\td \rho^2+ \td z^2),
\end{equation}
where $\partial_A = (k, m_i)$ for $(A=0, \dots D-3)$, $\rho^2 := -\det g_{AB}$ and $dz:=-\star_2 d\rho$, with $\star_2$ the Hodge dual on the orbit space $\hat{M}: = M/G$. Note that since $\partial_A$ are KVFs, the metric coefficients only depend on $\rho$ and $z$.

The orbit space $\hat{M}$ is a simply connected manifold with boundaries and corners, with the boundary given by the $\rho = 0$ axis in Weyl-Papapetrou coordinates \cite{Hollands:2007aj,Hollands:2008fm}. Furthermore the corners (rod points), occurring at specific values of $z$, divide this boundary up into intervals (rods) corresponding to either axes where an integer linear combination of the KVFs vanish or horizon orbit spaces. The presence of finite axis rods is generically associated with conical singularities: for a given axis rod $I$ with rod vector $v$, there is a conical singularity unless \cite{Harmark:2004rm}
\begin{equation}
\label{EM:conical}
    \lim_{\rho \to 0, \;  z\in  I} \frac{\rho^2 e^{2\nu}}{g(v,v)} =  1.
\end{equation}
Note that we are taking the angles $\phi^i$ to have a standard $2 \pi$ period. We will solve some of these conditions explicitly in Section \ref{sec:examples} when we consider black Saturn and black lens solutions.

Instead of looking at the $2$-dimensional $\hat{M}$, to better understand the hidden symmetries of this theory we must instead reduce to a $3$-dimensional base space which we will denote $M_3$. There are two obvious ways of doing this using the symmetries available, corresponding to either quotienting out the metric by $U(1)^{D-3}$ or $U(1)^{D-4} \times \mathbb{R}$. For the reduction by $U(1)^{D-3}$ one can write the metric as 
\begin{equation}
    g = e^{2 \nu}(\td \rho^2 + \td z^2) - \gamma^{-1}\rho^2 \td t^2 + \gamma_{ij}(\td \phi^i + w^i \td t)(\td \phi^j + w^j \td t),
\end{equation}
where $\gamma = \det \gamma_{ij}$ and we've taken $x^A = (t, \phi^i)$ and so $\partial_0 = k$ and $\partial_i = m_i$.

Next we introduce some potentials as follows: First define the electric and magnetic potentials $\Phi$ and $\Psi_i$ by 
\begin{equation}
\label{UUMaxwell}
    \td\Phi = \iota_{1}  \cdots \iota_{D-3} \star F, \qquad
    \td\Psi_i = \iota_i F,
\end{equation}
where $\iota_i := \iota_{m_i}$. $\Phi$ and $\Psi_i$ are well-defined (up to constants) through Maxwell's equations and the topological censorship theorem. Next we define the twist 1-forms $\Omega_i$ by
\begin{equation}
    \Omega_i = \star(m_1 \wedge \dots \wedge m_{D-3} \wedge \td m_i),
\end{equation}
where by abuse of notation we've used $m_i$ to stand in for their covector metric duals. Since $\iota_j\Omega_{i} = 0$, these can be viewed as 1-forms on $M_3$ or more explicitly
\begin{equation}
    \Omega_{i} = |\gamma|^{-1/2}\star_3 \td m_i
\end{equation}
where the $m_i$ are viewed as functions on $M_3$ and $\star_3$ is the Hodge dual on this space. By further reducing down to $\hat{M}$ (and using the fact that $\iota_0\Omega_{i} = 0$), $\Omega_i$ can be related to $w^i$ through
\begin{equation}
    \Omega_i = \rho^{-1} \gamma \gamma_{ij} \star_2 \td w^j,
\end{equation}
where $\star_2$ is the Hodge dual on $\hat{M}$

Defining the Levi-Civita connection $\tD$ on $M_3$, we can now write the Einstein-Maxwell equations for the Killing part of the metric and the potentials as\footnote{These equations are identical to those appearing in \cite{Ida:2003wv}, with a corrected factor of 2 on terms quadratic in the Maxwell potentials.}
\begin{equation}
\begin{aligned}
\label{EMgamma}
    \tD^2\gamma_{ij} = &\gamma^{kl}\tD \gamma_{ik}\cdot \tD \gamma_{jl} - \gamma^{-1} \Omega_i \cdot \Omega_j - 4\tD \Psi_i \cdot \tD\Psi_j \\
    &+ \frac{4}{D-2}\gamma_{ij}(\gamma^{kl}\tD\Psi_k \cdot \tD\Psi_l - \gamma^{-1}\tD\Phi \cdot \tD\Phi),
\end{aligned}
\end{equation}
\begin{equation}
\label{EMOmega}
    \tD \cdot \Omega_i = \gamma^{-1}\tD \gamma \cdot \Omega_i + \gamma^{jk} \tD \gamma_{ij} \cdot \Omega_k,
\end{equation} 
\begin{equation}
    \label{EMPhi}
    \tD^2\Phi = \gamma^{-1}\tD \gamma \cdot \tD\Phi + \gamma^{ij}\tD\Psi_i \cdot \Omega_j,
\end{equation}
\begin{equation}
    \label{EMPsi}
    \tD^2\Psi_i = \gamma^{jk} \tD \gamma_{ij} \cdot \tD\Psi_k - \gamma^{-1}\tD\Phi \cdot \Omega_i ,
\end{equation}
and
\begin{equation}
\label{EMOmegaTwist}
    \td \Omega_i = 4 \td \Phi \wedge \td \Psi_i.
\end{equation}
At first glance it may appear as though these equations involve the conformal factor $e^{2 \nu}$ through the inner product and connection on $M_3$, however this turns out not to be the case. This is due to the fact that i) $\Omega_i$ and the functions we are considering are invariant under the action of the stationary KVF $k$ and ii) the $(\rho,z)$ part of the metric is conformally flat. Combining these pieces of information we find that the inner product on $M_3$ acts like the inner product on $\mathbb{R}^3$ in cylindrical polars, up to a conformal factor which can be scaled away in the equations above.

The equations for $\nu$ come from gravity on $M_3$ coupled to $\gamma_{ij}$ and the various potentials. Reducing this to two dimensions gives a pair of PDEs
\begin{align}
\label{nuintegrability}
    \begin{split}
        \frac{1}{\rho}\left(\nu_{,\rho} + \frac{1}{2}\gamma^{-1}\gamma_{,\rho}\right) = X_{\rho \rho} - X_{zz}, \qquad
        \frac{1}{\rho}\left(\nu_{,z}+ \frac{1}{2}\gamma^{-1}\gamma_{,z}\right) =& 2 X_{\rho z},
    \end{split}
\end{align}
where
\begin{align}
\label{Xijnu}
    \begin{split}
        X_{pq} = \gamma^{-1}\Phi_{,p} \Phi_{,q} + \gamma^{ij} &\Psi_{i,p}\Psi_{j,q} + \frac{1}{8} \gamma^{-2} \gamma_{,p}\gamma_{,q} + \frac{1}{8} \gamma^{ij} \gamma^{kl} \gamma_{ik,p}\gamma_{jl,q}+ \frac{1}{4} \gamma^{-1}\gamma^{ij}(\iota_p \Omega_i) (\iota_q \Omega_j),
    \end{split}
\end{align}
and $p,q$ run over $\rho$ and $z$. The integrability of these equations can be established using (\ref{EMgamma}) to (\ref{EMOmegaTwist}).

We have been considering the reduction over just the axial KVFs i.e. over $U(1)^{D-3}$, however there is another obvious reduction one can perform over the stationary KVF and all but one of the axial KVFs i.e. over $U(1)^{D-4} \times \mathbb{R}$. Without loss of generality we can choose the leftover KVF to correspond to $\partial_{D-3}$. Then we can write the metric as
\begin{equation}
    g = e^{2 \nu}(\td \rho^2 + \td z^2) + h_{\mu \nu}(\td x^\mu + y^\mu \td \phi^{D-3})(\td x^\nu + y^\nu \td \phi^{D-3}) - h^{-1}\rho^2 (\td \phi^{D-3})^2,
\end{equation}
where $h = \det h_{\mu\nu}$ and $\mu, \nu = 0, \dots , D-4$. We can define potentials in a similar way to the other reduction:
\begin{equation}
\label{RUPots}
\begin{gathered}
    \td R = \iota_{0}  \cdots \iota_{D-4} \star F, \qquad
    \td S_\mu = \iota_\mu F, \\
    Z_\mu = \star(k \wedge m_1 \wedge \cdots \wedge m_{D-4} \wedge \td e_\mu),
\end{gathered}
\end{equation}
where $e_\mu = (k,m_{i\neq D-3})$. The Einstein-Maxwell equations ((\ref{EMgamma}) to (\ref{Xijnu})) then take the same form with $\{h_{\mu\nu},R,S_\mu,Z_\mu\}$ replacing $\{\gamma_{ij},\Phi,\Psi_i,\Omega_i\}$.

\section{Charging transformations}
\label{sec:transformations}
We now set $D=5$ and consider electrostatic solutions - we will shortly see how this condition can be written in terms of the potentials adapted to each reduction. We can then derive non-trivial transformations between solutions in this class by looking at the symmetries of the Einstein-Maxwell equations described in the previous section.

\subsection{Charging black holes}
\label{ChargingRed}
We will start with the slightly less natural reduction over $\mathbb{R} \times U(1)$. Without loss of generality we can take this $U(1)$ to be generated by $\partial_1$. Then under this reduction staticity implies that 
\begin{equation}
    h_{01} = h_{10} = 0, \qquad Z_0 = 0, \qquad y^0 = 0
\end{equation}
and a pure electric spacetime must also satisfy
\begin{equation}
    R = 0, \qquad S_{\mu\neq 0} = 0.
\end{equation}
For convenience we define $S = S_0$, then the metric can be written
\begin{equation}
\label{metricRed1}
    g = e^{2 \nu}(\td \rho^2 + \td z^2) - e^{2V_0} \td t^2 + e^{2V_1} (\td \phi^1 + y \td \phi^2)^2 + e^{2V_2} (\td \phi^2)^2,
\end{equation}
where $e^{2V_0} = h_{00}$, $e^{2V_1} = h_{11}$, $y = y^1$ and $V_0 + V_1 + V_2 = \ln \rho$. 

The Einstein-Maxwell equations for $V_0$ and $S$ come from (\ref{EMgamma}) and (\ref{EMPsi}) and are given by
\begin{equation}
    \label{V01}
    \tD^2 V_0 = \alpha^{-2} e^{-2V_0} (\tD S)^2, \qquad
	\tD^2 S = 2\tD V_0 \cdot \tD S
\end{equation}
where $\alpha = \frac{\sqrt{3}}{2}$. Note that these two equations only depend on $V_0$ and $S$ and so can be solved independently to the rest of the equations.

We now consider the target space $T$, defined by the equations for $V_0$ and $S$ (\ref{V01}). The equations can be viewed as coming from the Lagrangian
\begin{equation}
    \mathcal{L} = \tD V_0^2 - \alpha^{-2} e^{-2V_0}\tD S^2.
\end{equation}
Therefore by defining coordinates $X^\pm = e^{V_0} \pm \alpha^{-1} S$, we can write the metric on $T$ as
\begin{equation}
    ds^2 = 4\frac{dX^+dX^-}{(X^+ + X^-)^2},
\end{equation}
which we recognise as $\textbf{AdS}_2$ in lightcone coordinates. It has isometries given by
\begin{equation}
	\label{Gammasym}
		X^\pm \to a X^\pm, \qquad
		X^\pm \to X^\pm \pm b, \qquad
		X^\pm \to \frac{X^\pm}{1 \mp c X^\pm},
\end{equation}
for real constants $a$, $b$ and $c$, with the KVFs corresponding to these transformations generating the Lie algebra $sl(2,\mathbb{R})$. These are hidden symmetries of the original equations (\ref{V01}). The dilation and translation transformations are both trivial, corresponding to rescaling $t$ and gauge transforming $S$ respectively. However, the third transformation is more interesting and can be used to generate a non-trivial 1-parameter family of new solutions given a starting seed solution. 

It is convenient when performing the third transformation to simultaneously rescale $t$ and gauge transform $S$ in order to manifestly preserve the asymptotic conditions. Specifically we impose that if $e^{2V_0} \to 1$ and $S \to 0$ at asymptotic infinity for the seed metric then these conditions should hold for the final metric as well. Then the transformation can be written as
\begin{equation}
\label{V0STransforms}
	e^{2V_0} \to e^{2V_0} L^{-2}, \qquad
	S \to \frac{(1 - c \alpha^{-1} S)\left(S - \alpha c\right) + \alpha c e^{2V_0}}{(1 - c \alpha^{-1} S)^2 - c^2 e^{2V_0}},
\end{equation}
where
\begin{equation}
\label{LDef}
	L = \frac{(1 - c \alpha^{-1} S)^2 - c^2 e^{2V_0}}{1 - c^2}.
\end{equation}
The equations for the other metric components (\ref{EMgamma}), (\ref{nuintegrability}) imply that they transform as
\begin{equation}
\label{OtherTransforms1}
    e^{2V_i} \to e^{2V_i}L\quad (i=1,2), \qquad
    y \to y, \qquad
    e^{2\nu} \to e^{2\nu}L.
\end{equation}
Note that the condition $V_0 + V_1 + V_2 = \ln \rho$ is invariant under this transformation.
	
In order to preserve signature and avoid creating new singularities under this transformation, $L$ must be positive which is satisfied if and only if $-1<c<1$. To show that this implies $L>0$, it is sufficient to show that $X^{\pm} < 1$ \footnote{We are using the fact that $e^{2V_0} \to 1$ and $S \to 0$ at asymptotic infinity and we also assume that the mass $M$ and charge $Q$ obey $M > |Q|$.}, a result which is the content of lemma 2 in \cite{Kunduri:2017htl}. We also note that $g_{ij}$ transforms with an overall factor of $L$ meaning that a rod vector of the starting solution is a rod vector of the end solution and so the rod structure is partially preserved.

Since this transformation preserves asymptotic flatness whilst adding an electric field, it can be physically interpreted as adding electric charge into the bulk of the spacetime, or equivalently adding charge to black hole horizons that are present. In fact if one were to apply this to a higher dimensional Schwarzschild black hole, then the transformed metric would be a Reissner–Nordström black hole with charge proportional to $c$ - this is essentially guaranteed by the static uniqueness theorem for black hole spacetimes \cite{Gibbons:2002ju}.

\subsection{Immersing black holes in an electric background \label{VPsisection}}
Now we consider the reduction over $U(1)^2$. In this case staticity implies that 
\begin{equation}
    \Omega_i = 0, \qquad w^i = 0,
\end{equation}
and a pure electric spacetime must also satisfy
\begin{equation}
    \Psi_i = 0.
\end{equation}
The metric can be written in the form
\begin{equation}
    g = e^{2 \nu}(\td \rho^2 + \td z^2) - \rho^2 e^{-2W} \td t^2 + e^{2W_1}(d\phi^1 + u d\phi^2)^2 + e^{2W_2}(d\phi^2)^2,
\end{equation}
where $e^{2W_1} = \gamma_{11}$, $u = \gamma_{12} \gamma_{11}^{-1}$, $e^{2W_2} = \gamma_{22} - \gamma_{12}^2 \gamma_{11}^{-1}$ and $W := W_1 + W_2$.

The Einstein-Maxwell equations for $W$ and $\Phi$ (\ref{EMgamma}), (\ref{EMPhi}) are
\begin{align}
	\tD^2W = -\alpha^{-2} e^{-2W}(\tD\Phi)^2, \qquad \tD^2\Phi = 2\tD W \cdot \tD\Phi.
\end{align}
An almost identical analysis as in the previous section applies to these equations with the only difference being that in this case the target space $T \cong \textbf{H}^2$, the hyperbolic plane. The isometries of $T$ determine the transformations for $W$ and $\Psi$ as before. The non-trivial 1-parameter family of transformations is given by
\begin{equation}
\label{OtherTransforms2}
\begin{gathered}
	e^{2W} \to e^{2W} M^{-2}, \qquad
    e^{2W_i} \to e^{2W_i}M^{-1}, \qquad
    u \to u \\
	\Phi \to \left[\Phi (1 + k \alpha^{-1} \Phi) + \alpha k e^{2W} \right] M^{-1}, \qquad
    e^{2\nu} \to e^{2\nu}M^{2}
\end{gathered}
\end{equation}
where
\begin{equation}
\label{MDef}
	M = (1 + k \alpha^{-1}\Phi)^2 + k^2 e^{2W}
\end{equation}
and $k$ is a real parameter. Note that the condition $W - W_1 - W_2 = 0$ is invariant under this transformation.

Similarly to the previous transformation, $M$ must be positive to preserve signature and avoid creating new singularities. $M>0$ follows immediately from the definition and there are no restrictions on $k$. As before $g_{ij}$ transforms with just an overall factor ($M^{-1}$ in this case) and so the transformed solution has the same rod vectors as the seed.

An important difference between this transformation and the previous one is that now asymptotic flatness can no longer be preserved. In fact this transformation takes asymptotically flat spacetimes to asymptotically Melvin ones. These are spacetimes that asymptotically look like the 5d electric Melvin universe, a spacetime with metric
\begin{equation}
\label{Melvin}
    ds^2 = M^2 \frac{\mu}{\rho^2 + \mu^2}(\td \rho^2 + \td z^2) -M^2\td t^2 + M^{-1}\left(\mu(\td \phi^1)^2 + \frac{\rho^2}{\mu}(\td \phi^2)^2\right),
\end{equation}
where
\begin{equation}
    M = 1 + k^2 \rho^2, \qquad \mu = \sqrt{\rho^2 + z^2} - z,
\end{equation}
and $k$ determines the strength of the electric field. Therefore one can think of this transformation as taking a spacetime and immersing it in an electric background.

\subsection{Combined transformation}
\label{ssec:combined}
We now consider applying these two transformations consecutively to neutral, static, AF black hole spacetimes. By convention we will take $m_2$ to be the rod vector for the left semi-infinite rod $I_L$ (i.e. the rod with $z\to-\infty$) and $m_1$ for the right semi-infinite rod $I_R$ (i.e. the rod with $z\to\infty$). As discussed in the previous section, the first transformation will give the black holes an electric charge and then the second will immerse them in a background electric field. We know (from the static uniqueness theorem) that these static seed solutions will generally have some kind of singularity. If these are conical in nature we will show how tuning the parameters of these transformations might allow one to cure these singularities to give regular solutions.

Consider a static seed metric as in (\ref{metricRed1}), with $S = 0$, i.e. a neutral solution. If we charge this solution using the transformation associated to the $\mathbb{R} \times U(1)$ reduction (\ref{V0STransforms}), (\ref{OtherTransforms1}), (\ref{LDef}), we find that
\begin{equation}
    g = e^{2 \nu} L (\td \rho^2 + \td z^2) - e^{2V_0} L^{-2} \td t^2 + e^{2V_1} L (\td \phi^1 + y \td \phi^2)^2 + e^{2V_2} L (\td \phi^2)^2,
\end{equation}
\begin{equation}
\label{TransS}
    S = \frac{\alpha c (e^{2V_0} - 1)}{1-c^2e^{2V_0}}
\end{equation}
where
\begin{equation}
    L = \frac{1-c^2e^{2V_0}}{1-c^2}.
\end{equation}

Next we need to convert these into the variables adapted to the $U(1)^2$ reduction. This is trivial for the metric components
\begin{equation}
\label{TransMetric}
    e^{2W^{(0)}} = \rho^2 e^{-2V_0}L^2, \qquad e^{2W_1^{(0)}} = e^{2V_1} L, \qquad  u = y, \qquad e^{2W_2^{(0)}} = e^{2V_2} L,
\end{equation}
where we use $(0)$ superscripts for this intermediate solution for later convenience. To do something similar for the Maxwell potential, we first work from the definition of $S$ (\ref{RUPots}) and $\Phi^{(0)}$ (\ref{UUMaxwell}) (and use the fact that $F$ is purely electric) to find that
\begin{equation}
    \td \Phi^{(0)} = \rho^{-1} e^{2W^{(0)}} \star_2 \td S.
\end{equation}
Using the expression for $e^{2W^{(0)}}$ (\ref{TransMetric}) and $S$ (\ref{TransS}) in terms of $V_0$ this simplifies to
\begin{equation}
\label{PhiV0}
    \td \Phi^{(0)} = \frac{2 \alpha c}{1-c^2} \rho \star_2 \td V_0.
\end{equation}

$V_0$ is an axially symmetric harmonic function on $\mathbb{R}^3$, which can be seen from (\ref{V01}) since the seed is neutral (or even just considering the above equation on $M_3$ and acting with the exterior derivative on both sides). We also know that $V_0$ must tend to $0$ at asymptotic infinity (since the solution is AF) and be smooth everywhere except for on horizon rods where it should diverge as $\ln \rho$ (this is necessary for a smooth horizon). A candidate form for $V_0$ that satisfies these constraints can be written as 
\begin{equation}
\label{V0form}
    V_0 = \frac{1}{2} \sum_{H} \ln \frac{\mu_{H-1}}{\mu_H},
\end{equation}
where the sum is over all horizon rods $H = (z_{H-1},z_H)$ and 
\begin{equation}
\label{muDef}
    \mu_k = \sqrt{\rho^2 + (z-z_k)^2} - (z-z_k).
\end{equation}
Note that $\ln \mu_k$ are axially symmetric harmonic functions and $\frac{1}{2}\ln \left(\mu_{H-1}/\mu_H\right)$ is smooth everywhere apart from $\rho=0, z \in H$ where it diverges as $\ln \rho$ as the $z$-axis is approached. Considering another function $V_0'$ satisfying these constraints, it is simple to see that $V_0'-V_0$ is a smooth and bounded harmonic function and so must be constant everywhere \footnote{A more complete analysis would be needed to show that $V_0'-V_0$ is also free from divergences at the endpoints of horizon rods.}. This gives justification that the form of $V_0$ given above is unique up to a rescaling of the $t$ coordinate.

Combining (\ref{V0form}) with the identity 
\begin{equation}
    \rho \star_2 d \ln \mu_k = - d \bar{\mu}_k,
\end{equation}
where
\begin{equation}
\label{mubarDef}
    \Bar{\mu}_k = \rho^2/\mu_k = \sqrt{\rho^2 + (z-z_k)^2} + (z-z_k),
\end{equation}
and using the result in the equation for $\Phi^{(0)}$ in terms of $V_0$ (\ref{PhiV0}), we finally get a solution for $\Phi^{(0)}$
\begin{equation}
\label{Phi0Def}
    \Phi^{(0)} = -\frac{\alpha c}{1-c^2} \sum_{H} (\Bar{\mu}_{H-1}-\Bar{\mu}_{H}),
\end{equation}
which is valid up to an arbitrary additive constant. Notice that we have chosen a gauge for $\Phi^{(0)}$ such that $\Phi^{(0)}|_{I_L} = 0$

Now we can use the transformation associated to the $U(1)^2$ reduction (\ref{OtherTransforms2}), (\ref{MDef}) on this charged solution to give
\begin{equation}
\label{gfinal}
    g = e^{2 \nu} L M^2 (\td \rho^2 + \td z^2) - e^{2V_0} L^{-2} M^2 \td t^2 + e^{2V_1} L M^{-1} (\td \phi^1 + y \td \phi^2)^2 + e^{2V_2} L M^{-1} (\td \phi^2)^2,
\end{equation}
\begin{equation}
\label{PhiTransformed}
    \Phi = \left[\Phi^{(0)}(1 + k \alpha^{-1} \Phi^{(0)}) + \alpha k e^{2 W^{(0)}}\right]
\end{equation}
where
\begin{equation}
    M = (1 + k \alpha^{-1} \Phi^{(0)})^2 + k^2 e^{2 W^{(0)}}
\end{equation}
and $e^{2W^{(0)}}, \Phi^{(0)}$ are given in (\ref{TransMetric}), (\ref{Phi0Def}). We finally note that since $M$ and $L$ are smooth in $\rho^2$ and have non-zero limits on horizon rods, if the seed solution has regular horizons then so too will the transformed solution.

\subsubsection{Conical singularities}
\label{ssec:conical}
Lastly we discuss conical singularities. For an axis rod $I_a$ with rod vector $v_a$, there is a conical singularity unless the balance condition (\ref{EM:conical}) is satisfied (taking $\phi^i$ to have period $2 \pi$). We can impose that these conditions are automatically satisfied for the left and right semi-infinite rods by appropriately rescaling the angles $\phi^i$. After the combined transformation the expression on the LHS of (\ref{EM:conical}) will pick up a factor of $(M^a_0)^3$ where we define
\begin{equation}
\label{M0aDef}
    M^a_0 = \lim_{\rho \to 0, \;  z\in  I_a} M = (1 + k \alpha^{-1} \Phi^{(0)}|_{I_a})^2,
\end{equation}
and we have used the fact that $e^{2W^{(0)}}$ vanishes on axis rods. From our solution for $\Phi^{(0)}$ (\ref{Phi0Def}) and the limiting behaviour of $\bar{\mu}_k$ on axis rod $I_a$, we find that
\begin{equation}
\label{Phi0a}
    \Phi^{(0)}|_{I_a} = -\frac{2 \alpha c}{1-c^2}\sum_{H<I_a}\ell_H,
\end{equation}
where the sum is over all horizon rods $H=(z_{H-1},z_H)$ less than the axis rod $I_a$ (in the sense of intervals on the $z$-axis) and $\ell_H = z_H-z_{H-1}$ is the rod length of the horizon $H$. Note that this expression is constant on $I_a$, which can alternately be seen directly from the definition of $\Phi$ (\ref{UUMaxwell}) using the fact that the rod vector $v_a=0$ on axis rod $I_a$. Next from the definition of $M^a_0$ above (\ref{M0aDef}), we see that
\begin{equation}
\label{M0a}
    M^a_0 = \left(1 -\frac{2 k c}{1-c^2} \sum_{H<I_a}\ell_H\right)^2.
\end{equation}

Consider the left axis rod $I_L$. Then we see from (\ref{M0a}) that $M^L_0 = 1$ meaning that there is no conical singularity in the transformed metric. On the other hand on the right rod $I_R$, we have
\begin{equation}
    M^R_0 = \left(1 -\frac{2 k c}{1-c^2} \sum_{H}\ell_H\right)^2
\end{equation}
where the sum is over all horizon rods $H$. Therefore the left hand side of (\ref{EM:conical}) is given by
\begin{equation}
\label{NDef}
    N^2 :=  (M^R_0)^3 = \left(1 -\frac{2 k c}{1-c^2} \sum_H \ell_H \right)^6,
\end{equation}
for the transformed metric. We see that $N^2$ is not equal to $1$ unless either $c,k=0$ or there are no horizons (we return to these cases below). Otherwise if we want to remove the conical singularity on $I_R$ we must either relax the assumption that $\phi^1$ has a period of $2\pi$ or equivalently rescale $\phi^1$. We take the second option and rescale $\phi^1 \to N \phi^1$, assuming $N>0$ for convenience. This puts the metric into the new form
\begin{equation}
\label{finalg}
\begin{aligned}
    g =&\; e^{2 \nu} L M^2 (\td \rho^2 + \td z^2) 
    \\&- e^{2V_0} L^{-2} M^2 N^{-2} \td t^2 + e^{2V_1} L M^{-1} (N \td \phi^1 + y \td \phi^2)^2 + e^{2V_2} L M^{-1} (\td \phi^2)^2,
\end{aligned}
\end{equation}
where we have also taken $t \to N^{-1} t$ in order to maintain the $W - W_1 - W_2 = 0$ condition. An immediate consequence of this is that $\partial_1$ transforms as $\partial_1 \to N^{-1} \partial_1$ under this coordinate change and so a rod vector should transform as well i.e. as $v = p \partial_1 + q \partial_2 \to N^{-1} p \partial_1 + q \partial_2$ for constants $p$ and $q$. This is compatible with the earlier statements that these charging transformations shouldn't change rod vectors since all that is changing is the coordinates used to describe them. For the orbits of this rod vector to be closed we now have the requirement that either $p=0$ or $q N p^{-1}$ is rational. We will return to this condition when we discuss the black lens spacetime in the next section.

We now consider the special cases where $N^2 = 1$ which we ignored above. First consider $c=0$. This implies that $\Phi^{(0)} = 0$ and so $N^2 = M_0^a = 1$ for all axis rods $I_a$ which in turn implies that the transformation doesn't affect the balance conditions (\ref{EM:conical}). Similarly when $k=0$, the conditions are again unaffected by the transformation. We therefore see that both the charging and immersing transformations are needed to act non-trivially in order to have a chance at removing singularities. Lastly, consider a soliton solution, a spacetime with no black hole horizons. In this case $e^{2V_0} = 1$ and so $S=0, L=1$, which means that the transformation is independent of $c$ - using the same arguments as above this immediately implies that the transformation cannot be used to cure conical singularities in the transformed solution.

Analysis of the conical singularity condition for the finite axis rods is difficult to do in general, although in the case where the seed is a Weyl solution \cite{Emparan:2001wk}, some progress can be made, see Appendix \ref{ap:Weyl} for details. 

\section{Examples}
\label{sec:examples}
We now consider applying the combined charging transformation of the previous section to some neutral seed solutions. For flat space, the transformation just gives the five-dimensional electric Melvin universe (\ref{Melvin}) with dependence on the parameter $c$ dropping out as described in the previous section since there are no horizons. If we instead consider charging the five-dimensional Schwarzschild solution then we find a Reissner–Nordström black hole in a background field - this must be the case from uniqueness. In both cases the charged solutions are automatically regular since neither solution has finite axis rods which could have conical singularities. 

The simplest seed that the charging transformation could help balance is the static black ring. This solution has a single finite axis rod with an associated conical singularity (see Figure \ref{fig:staticBR} for it's rod structure). Charging this solution introduces two new parameters $c,k$ which one can tune in order to remove the singularity and find a regular ring in a background field. We will not demonstrate this balancing here explicitly since we next consider the black Saturn solution from which the ring can be found as a particular limit.

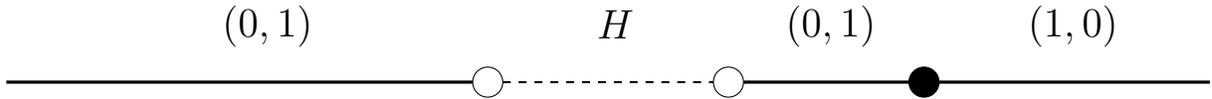
\BRfig{fig:staticBR}{The rod structure for a black ring solution. The solid lines represent axis rods with rod vectors written above them in terms of the $(\partial_1,\partial_2)$ basis. The dashed line corresponds to a horizon $H$. Solid circles denote corners between two axis rods and empty circles denote corners between an axis rod and a horizon rod.}

\subsection{Black Saturn \label{BS}}
The neutral static black Saturn solution can be constructed from its rod structure (Figure \ref{fig:staticBS}) as a Weyl solution \cite{Emparan:2001wk}, with its metric given by 
\begin{equation}
\begin{gathered}
	e^{2V_0} = \frac{\mu_1\mu_3}{\mu_4\mu_2}, \qquad
	e^{2V_1} = \mu_4, \qquad
	e^{2V_2} = \rho^2\frac{\mu_2}{\mu_1\mu_3},\\
	e^{2\nu} = \mu_4  \frac{r_{12}^2 r_{23}^2 r_{14} r_{34}}{r_{13}^2 r_{24}\prod_{i=1}^{4}r_{ii}^2},
\end{gathered}
\end{equation}
where
\begin{equation}
    r_{kl} = \rho^2 + \mu_k \mu_l,
\end{equation}
$\mu_k$ is given by (\ref{muDef}) and the rod points obey $z_1<z_2<z_3<z_4$. 

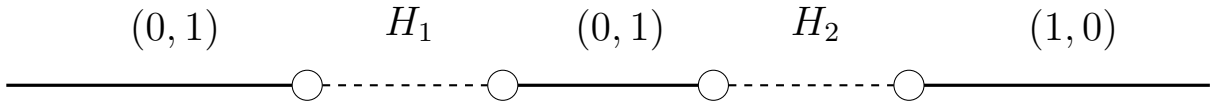
\BSfig{fig:staticBS}{The rod structure for a black Saturn solution. Horizon $H_1$ corresponds to a black ring and $H_2$ to an $S^3$ black hole.}

The metric physically corresponds to an $S^3$ black hole surrounded by a black ring in a flat background. One can isolate the central $S^3$ black hole by taking $z_2 \to z_1$, essentially removing the black ring horizon rod (note that this also causes the dependence of the metric on $z_1$ to drop out). There is also a limit to the black ring by taking $z_4 \to z_3$ which removes the $S^3$ horizon. Since this is a static, AF solution to vacuum gravity (which is neither flat space nor a Schwarzschild black hole), it cannot be a smooth solution \cite{Gibbons:2002ju}. As expected this is because of a conical singularity on the finite axis rod $I_C = (z_2,z_3)$ which has rod vector $v_C = m_2$. Explicitly we find that
\begin{equation}
\label{BSvacConical}
\lim_{\rho \to 0,\;  z \in  I_C} \left(\frac{\rho^2 e^{2\nu}}{e^{2V_2}}\right) = \frac{(z_3-z_2)^2(z_4-z_1)}{(z_3 - z_1)^2(z_4 - z_2)},
\end{equation}
is always less than $1$ and so the conical singularity cannot be removed through tuning $z_k$ alone (see (\ref{EM:conical})). This expression can also be determined straightforwardly from the general case presented in Appendix \ref{ap:Weyl}, see (\ref{xaWeyl}).

We now charge this solution in the way described in the previous section. The equations for the metric components and $\Phi$ are trivially given from the general formalism  (\ref{finalg}), (\ref{PhiTransformed}). Expression (\ref{BSvacConical}) picks up a factor of $(M^C_0)^3$ (\ref{M0aDef}) where $M^C_0$ is given by
\begin{equation}
    M^C_0 = \left(1 -\frac{2 k c (z_2-z_1)}{1-c^2}\right)^2,
\end{equation}
using (\ref{M0a}). This means that the new balance condition can now be solved for $k$ to give
\begin{equation}
\label{BSksol}
    k = \frac{1-c^2}{2c(z_2-z_1)} \left(1 \pm \left[\frac{z_3 - z_1}{z_3-z_2}\left(\frac{z_4 - z_2}{z_4-z_1}\right)^{1/2}\right]^{1/3}\right),
\end{equation}
where we are assuming $c \neq 0$. This gives two disjoint families of regular solutions with a stronger external field for the positive sign and a weaker external field for the negative sign. Setting $z_4=z_3$ gives a balanced static charged black ring immersed in an electric background which matches the solution found in \cite{Kunduri:2004da} (for a choice of lower sign).

\subsection{Black hole with a "bubble"}
The black Saturn and black ring solutions have a single finite axis rod and are the only non-solitonic Weyl solutions with this property. We now consider a Weyl solution with two axis rods given by the rod structure in Figure \ref{fig:BHWB}, with metric
\begin{equation}
\begin{gathered}
	e^{2V_0} = \frac{\mu_1}{\mu_2}, \qquad
	e^{2V_1} = \frac{\mu_2\mu_4}{\mu_3}, \qquad
	e^{2V_2} = \rho^2\frac{\mu_3}{\mu_1\mu_4},\\
	e^{2\nu} = \frac{r_{12}^2r_{23} r_{24} r_{34}}{r_{13} r_{14} \prod_{i=1}^{4}r_{ii}^2},
\end{gathered}
\end{equation}
where $z_1<z_2<z_3<z_4$. This solution represents an $S^3$ black hole with a non-trivial $2$-cycle (or "bubble") in the DOC. In particular the finite axis rod $I_2$ lifts to an $S^2$ and the finite axis rod $I_3$ lifts to a non-contractible $2-$disk in a constant time slice of the full spacetime.

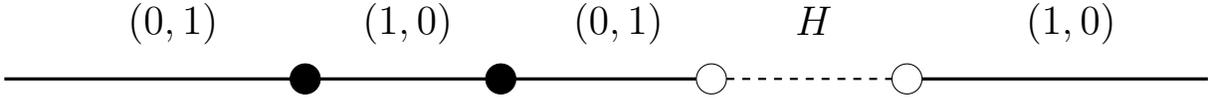
\BHWBfig{fig:BHWB}{The rod structure for a black hole with a bubble in the DOC.}

This solution has a conical singularity associated to each finite axis rod. We will concentrate on the the conical singularity for the finite axis rod $I_3$ and define
\begin{equation}
\label{x3bubble}
    x_3 = 
    \lim_{\rho \to 0, \;  z\in  (z_2,z_3)} \frac{\rho^2 e^{2\nu}}{e^{2V_2}} = \frac{(z_3-z_2)(z_4-z_2)}{(z_3-z_1)(z_4-z_1)}.
\end{equation}
Then the balance condition is $x_3 = 1$ (\ref{EM:conical}) which is never satisfied because $x_3<1$ as a result of the inequalities on the rod points $z_k$. After the transformation $x_3$ picks up a factor of $(M_0^3)^3$ (\ref{M0aDef}) in the balance condition and so we then have 
\begin{equation}
    (M_0^3)^3 x_3 = 1.
\end{equation}
However $M_0^3 = M_0^L = 1$ since there are no horizons between the first and third rod. Therefore there are no solutions to this equation and we cannot balance the charged black hole with bubble solution. 

There is another Weyl solution with a single horizon and two axis rods - its rod structure is represented in Figure \ref{fig:BHWDs}. This is again an $S^3$ black hole solution with now both of its two finite axis rods $I_2$ and $I_4$ lifting to non-contractible $2-$disks in a constant time slice of the full spacetime. The analysis of conical singularities here is very similar to the previous case and so we omit some details. We can define $x_2$ and $x_4$ similarly to (\ref{x3bubble}) as the LHS of (\ref{EM:conical}) for rods $I_2$ and $I_4$. After charging the solution we want to impose the new balance conditions
\begin{equation}
    \left(\frac{M_0^2}{M_0^R}\right)^{3}x_2 = 1,\qquad \left(M_0^4\right)^{3}x_4 = 1.
\end{equation}
Note that $M_0^2 = M_0^L = 1$ and $M_0^4 = M_0^R$ implying that $x_2 x_4 = 1$. One can also show that $x_2, x_4 < 1$ using the explicit form of the metric similarly to the previous example. From this we immediately see that there are no solutions to these equations implying that the charging transformation does not allow one to remove all the conical singularities of the seed solution.

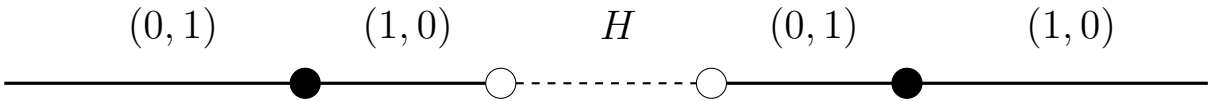
\BHWDsfig{fig:BHWDs}{The rod structure for another Weyl solution with a single horizon two axis rods.}

\subsection{Black lens \label{BL}}
We next consider a neutral static $L(n,1)$ black lens, a black hole spacetime with $L(n,1)$ lens space horizon topology. We use the metric given in \cite{Chen:2008fa}\footnote{The metric was originally derived in \cite{Ford:2007th} but written in Weyl coordinates and not recognised as describing a black lens spacetime. We have also used $\nu$ and $R$ in place of $c$ and $\kappa$ in \cite{Chen:2008fa} to avoid confusion with our charging parameters, and used $(\phi^1,\phi^2)$ in place of their $(\psi,\phi)$.} which is written in $(x,y)$ coordinates as 
\begin{equation}
\begin{aligned}
    g = - \frac{1+\nu y}{1+\nu x}\td t^2 + &\frac{2 R^2 (1+\nu x)}{(1-a^2)(x-y)^2 H(x,y)} \Bigg[\frac{H(x,y)^2}{1-\nu }\left(\frac{\td x^2}{G(x)} - \frac{\td y^2}{G(y)}\right)\\
    &+ (1-x^2)[(1-\nu-a^2(1+\nu y))\td \phi^2 - a\nu(1+y)\td \phi^1]^2\\
    &- (1-y^2)[(1-\nu-a^2(1+\nu x))\td\phi^1 - a\nu(1+x)\td \phi^2]^2\Bigg],
\end{aligned}
\end{equation}
where
\begin{equation}
    G(\zeta) = (1-\zeta^2)(1+\nu \zeta), \qquad H(x,y) = (1-\nu )^2 - a^2(1+\nu x)(1+\nu y).
\end{equation}
The constants lie in the ranges $0<\nu<1,-1<a<1$ and $R>0$ with the coordinates $(x,y)$ constrained by $-1\leq x \leq 1$ and $-1/\nu <y\leq -1$. We can see that this is not a Weyl solution in general since since $d/\phi^i$ are not generically hypersurface orthogonal. The rod structure is given in Figure \ref{fig:BLRSEM}. The left axis rod corresponds to $x=-1$, the horizon rod corresponds to $y=-1/\nu$, the finite axis rod corresponds to $x=1$ and the right axis rod corresponds to $y=-1$. The rod vector for the finite axis rod $I_D$ is given by $v_D = \partial_1 + n \partial_2$ with $n$ given by
\begin{equation}
\label{nDef}
    n = \frac{2a \nu }{1-\nu -a^2(1 +\nu )}.
\end{equation}
Requiring that the orbits of $v_D$ are closed imposes that $n$ is an integer, though we shall replace this condition with a slightly different one shortly  when we discuss the transformed solution. 

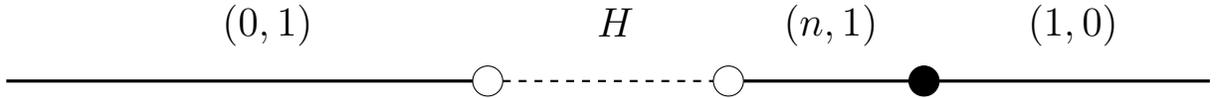
\BLfig{fig:BLRSEM}{The rod structure for a simple $L(n,1)$ black lens.}

The metric has a limit to a black ring by taking $n\to0$ (equivalently $a\to0$). Similarly there is a limit to a Schwarzschild black hole by taking $n\to\infty$ $\bigg($equivalently $a \to \pm \sqrt{\frac{1-\nu }{1+\nu }}\bigg)$. Again, as with the black Saturn solution, there is a conical singularity associated with the finite axis rod $I_D$ where $x=1$.

The metric of the spacetime after performing both the transformations we've discussed (\ref{finalg}), ({\ref{PhiTransformed})} can be written as
\begin{equation}
\label{gBLtransformed}
\begin{aligned}
    g = - \frac{1+\nu y}{1+\nu x} M^2 L^{-2} N^{-2} \td t^2 + &\frac{2 R^2 (1+\nu x)}{(1-a^2)(x-y)^2 H(x,y)}M^{-1}L \Bigg[\frac{H(x,y)^2}{1-\nu}M^3\left(\frac{\td x^2}{G(x)} - \frac{\td y^2}{G(y)}\right)\\
    &+ (1-x^2)[(1-\nu-a^2(1+\nu y)) \td \phi^2 - a\nu(1+y) N \td \phi^1]^2\\
    &- (1-y^2)[(1-\nu-a^2(1+\nu x)) N \td \phi^1 - a\nu(1+x)\td \phi^2]^2\Bigg],
\end{aligned}
\end{equation}
with $L,M,N$ given by (\ref{LDef}), (\ref{MDef}), (\ref{NDef}) and $\Phi$ given by (\ref{PhiTransformed}). Note that 
\begin{equation}
    \rho^2 = -\frac{4 R^2}{(x-y)^2} G(x) G(y)
\end{equation}
from the definition of $\rho$ in terms of the determinant of the Killing part of the metric. We also see that the rod vector for $I_D$ is now written $v_D = \partial_1 + \Bar{n} \partial_2$ for 
\begin{equation}
    \Bar{n} = N n = \left(1 -\frac{4 c k \nu R^2}{1-c^2} \right)^3 \frac{2a \nu }{1-\nu -a^2(1 +\nu )}.
\end{equation}
This means that we should now take $\Bar{n}$ to be an integer (and relax that requirement on $n$) to ensure that $v_D$ has compact orbits, giving a $L(\bar{n},1)$ black lens. 

Now we consider possible conical singularities on the axis rods. By construction there are no conical singularities on the semi-infinite axis rods as long as $\phi^i$ have periods $2\pi$. In order to cure the conical singularity condition for the finite axis rod $I_D$, we use the fact that the conformal factor for the neutral seed is given by
\begin{equation}
\begin{aligned}
    e^{2\nu} = &-\frac{2(1+\nu x)(x-y)H(x,y)}{(1-\nu)(1-a^2)R^2}\\
    &\left[(2 +\nu(1+x)+\nu(1-x)y)(\nu + x+ y + x y)(2 - \nu (1 -x - y -xy))\right]^{-1}.
\end{aligned}
\end{equation}
Therefore for the seed solution we find that
\begin{equation}
\label{BLseedconical}
   \lim_{x \to 1} \frac{\rho^2 e^{2\nu}}{g(v_D,v_D)} = \frac{(a^2(1+\nu) - (1-\nu) )^2} {(1-a^2)^2(1-\nu^2)}.
\end{equation}
The right hand side of this expression is less than $1$ for all allowed values of $a,\nu$ and therefore the balance condition (\ref{EM:conical}) cannot be satisfied in the neutral case (as expected).

As discussed in Section \ref{ssec:conical}, we know that under the combined charging transformation (\ref{BLseedconical}) will just be multiplied by an overall factor of $(M^D_0)^3$. $M^D_0$ is given by
\begin{equation}
    M^D_0 = \left(1 -\frac{4 k c \nu R^2}{1-c^2}\right)^2,
\end{equation}
where we have used (\ref{M0a}) and the fact that the horizon rod length is $z_2 - z_1 = 2 \nu R^2$ (see e.g. \cite{Chen:2008fa}). Therefore we can solve the balance condition (\ref{EM:conical}) for $k$ ($ k\neq 0$) to find
\begin{equation}
    k = \frac{1-c^2}{4c \nu R^2}\left(1\pm\left[\frac{(1-a^2)(1-\nu^2)^{1/2}}{|a^2(1+\nu) - (1-\nu) |}\right]^{1/3}\right).
\end{equation}
As with the black Saturn solution (\ref{BSksol}) this gives two distinct families of solutions corresponding to the charges of the transformations either having the same or opposite sign. Combining this with the expression for $\bar{n}$ we find
\begin{equation}
    \bar{n} = s\frac{2 a \nu (1-a^2)(1-\nu^2)^{1/2}}{((1-\nu)-a^2(1+\nu))^2},
\end{equation}
where $s$ gives the sign of $((1-\nu)-a^2(1+\nu))$. Any integer $\bar{n}$ can be found for suitable $a, \nu$, just as in the vacuum case with $n$.

\section{Discussion}
\label{sec:disc}
In this paper we have considered multi-axisymmetric, stationary solutions to $D$-dimensional Einstein-Maxwell gravity. The Einstein-Maxwell equations take a different form depending on whether one reduces to a three-dimensional base space over $\mathbb{R} \times U(1)^{D-4}$ or $U(1)^{D-3}$. 
After restricting to the $D=5$ electrostatic case we used these two different formulations to derive two distinct 1-parameter families of transformations. The first of these transformations preserves asymptotic flatness and can be interpreted as adding charge to black holes in a spacetime. The second transformation does not preserve asymptotic flatness but instead immerses the black hole in an external electric background known as the Melvin universe. Although the transformed solutions are no longer AF, they still preserve some of that structure, namely having an $S^3$ topology spatial cross-section at infinity. We illustrated these charging transformations by acting on a number of neutral singular seed solutions and attempted to tune the charging parameters to remove these singularities. The regular black lens we derived is of particular interest as it marks the first known example of a non-trivial regular black hole with lens space topology in Einstein-Maxwell gravity.

One avenue we have not fully explored in this paper is the question of reducing the Einstein-Maxwell equations to $3$ dimensions over other choices of KVFs. We have seen that each of the reductions we did consider gave a different perspective on the structure of the Einstein-Maxwell equations, making certain symmetries manifest. It would be interesting to see what structure would be exposed by a null reduction or more exotically some reduction involving the corotating KVF associated to some black hole (see \cite{Kunduri:2018qqt} for an example of this). Another interesting extension would be to determine all Weyl seeds that could be balanced using the combined transformation that we've discussed. Appendix \ref{ap:Weyl} presents the equations that must be satisfied in order to balance a transformed Weyl seed (\ref{WeylBal}), though we make no attempt at solving it in general. A plausible conjecture based off the limited examples considered is that only solutions with a single finite axis rod can be balanced using some appropriate charging transformation - this would imply that the black Saturn and black ring are the only Weyl solutions that can be balanced in this way. 

\vspace{0.1in}
 
\noindent {\bf Acknowledgements}. This work was funded by an EPSRC studentship and a Maxwell Institute Research Fellowship. I would like to thank James Lucietti for proposing this project and many helpful discussions.

\appendix
\section{Weyl solution seeds}
\label{ap:Weyl}
We consider charging a general Weyl solution seed using the methods of Section \ref{sec:transformations}. A five-dimensional Weyl solution is an electrostatic solution which has $2$ orthogonal commuting axial KVFs. In this case we can choose coordinates such that the Einstein-Maxwell equations for the Killing part of the metric (\ref{EMgamma}) simplify dramatically, giving $3$ Laplace equations for the metric coefficients \cite{Emparan:2001wk}. Writing the metric in the form (\ref{metricRed1}), we can set $y=0$ and we see that $\nabla^2 V_\mu = 0$ where $\nabla$ is the flat connection on $\mathbb{R}^3$. 

Consider a solution with $n+1$ rods $(-\infty,z_1), (z_1,z_2), \dots , (z_n,\infty)$. Without loss of generality we can fix the left rod to have rod vector $v_L = \partial_2$ and the right rod to have rod vector $v_R = \partial_1$. Then all the rod vectors $v_a$ must be equal to either $\partial_1$ or $\partial_2$ and we can write $V_\mu$ as \cite{Emparan:2001wk}
\begin{equation}
    V_\mu = v_{1\;\mu} \ln \rho  + \frac{1}{2}\sum_{k=1}^n (v_{k+1\;\mu} - v_{k\;\mu}) \ln \mu_k
\end{equation}
where $v_{a\; \mu}$ gives the $\mu$ component of the rod vector associated to some rod $I_a$ in the basis $(\partial_0,\partial_1,\partial_2)$. Note that for $V_0$ this reproduces the expression we had previously for a general static solution (\ref{V0form}). Next we can solve (\ref{nuintegrability}) for the conformal factor $e^{2\nu}$ to find \cite{Iguchi:2007zz}
\begin{equation}
    \nu = V_1 - \frac{1}{4} \sum_{k,l = 1}^n (v_{k+1\;\mu} - v_{k\;\mu})(v_{l+1\;\mu} - v_{l\;\mu})r_{kl}
\end{equation}
where $\mu_k$ is defined in (\ref{muDef}) and $r_{kl}$ is defined as 
\begin{equation}
    r_{kl} = \rho^2 + \mu_k \mu_l.
\end{equation}
There is an implicit sum over pairs of covariant $\mu$ indices which illustrates the fact that $\nu$ does not transform as a scalar. We now have the full metric written in terms of information from the rod structure. Note that this is the only Weyl solution given this rod structure as a result of the uniqueness theorem for (potentially) conically singular solutions \cite{Alaee:2019qhj}.

Next consider conical singularities of Weyl solutions. Define $x_a$ as the square of the LHS of (\ref{EM:conical}) for axis rod $I_a$, 
\begin{equation}
    x_a = \lim_{\rho \to 0, \;  z\in  I_a} \frac{\rho^2 e^{2\nu}}{g(v_a,v_a)}.
\end{equation}
Using our expressions for $V_\mu$ and $\nu$ above one can show that $x_L = x_R = 1$ and for finite axis rods $I_a$ we find that
\begin{equation}
\label{xaWeyl}
    \ln x_a = - \sum_{A=1}^{a-1}\sum_{M=a}^n (v_{A+1\;\mu} - v_{A\;\mu})(v_{M+1\;\mu} - v_{M\;\mu}) \ln(z_M - z_A).
\end{equation}
We know that $x_a \neq 1$ for at least some finite axis rods making the solution conically singular since the five-dimensional Reissner–Nordström solution is known to be the unique regular static solution \cite{Gibbons:2002bh,Gibbons:2002ju,Kunduri:2017htl}\footnote{In fact it appears that $x_a < 1$ for all finite axis rods for any rod structure considered, though we do not have a proof of this result.}

Now consider the charging transformations acting on this Weyl seed as in Section \ref{ssec:combined}. Then $V_\mu$ and $\nu$ will transform as encoded in (\ref{gfinal}).  The balance condition $x_a = 1$ for each finite axis rod $I_a$ will also pick up factors of $M_0^a$ and $M_0^R$ in the following way
\begin{equation}
\label{WeylBal}
\begin{gathered}
    (M_0^a)^3 x_a = 1, \qquad v_a = v_L,\\
    \left(\frac{M_0^a}{M_0^R}\right)^3 x_a = 1 , \qquad v_a = v_R.
\end{gathered}
\end{equation}
These give a complicated system of polynomials relating the finite rod lengths and charging parameters $k,c$. For a seed with a given rod structure, if these equations are consistent then this means that after charging the seed one can tune various parameters to give a regular solution. We illustrate this in Section \ref{sec:examples} with the examples of the black Saturn and two examples of black holes with non-trivial $2$-cylces in the DOC.

\printbibliography

\end{document}

%% file: Preamble.tex
\usepackage[utf8]{inputenc}
\usepackage{amssymb, amsmath, amsthm}
\usepackage[left=2cm,right=2cm,top=2cm,bottom=2cm]{geometry}
\usepackage{comment}
\usepackage[colorlinks=true,linktocpage=true,allcolors=blue]{hyperref}
\usepackage{appendix}
\usepackage{mathrsfs}

\usepackage{graphicx,subfig,tikz}
\usetikzlibrary{decorations.pathmorphing}

\usepackage{csquotes}
\MakeOuterQuote{"}

\usepackage[backend=bibtex, style=numeric, sorting=none, isbn=false, url = false, date = year, giveninits = true]{biblatex}
\newcommand{\tD}{\text{D}}
\newcommand{\td}{\text{d}}

\def\be{\begin{equation}}
\def\ee{\end{equation}}
\def\bea{\begin{eqnarray}}
\def\eea{\end{eqnarray}}

\include{TikzCode}

%% file: TikzCode.tex
\newcommand{\BRfig}[2]{\begin{figure}[h]
\centering
\subfloat{
\begin{tikzpicture}[scale=2, every node/.style={scale=1.3}]
\draw[very thick](-4,0)--(-0.9,0)node[black,left=2.1cm,above=.2cm]{$(0,1)$};
\draw[thick,dashed](-.7,0)--(.7,0)node[black,left=1.0cm,above=.3cm]{$H$};
\draw[very thick](0.9,0)--(2.0,0)node[black,left=.8cm,above=.2cm]{$(0,1)$};
\draw[very thick](2.2,0)--(4,0)node[black,left=1.4cm,above=.2cm]{$(1,0)$};
\draw[fill=white] (-.8,0) circle [radius=.1] node[black,font=\large,below=.1cm]{};
\draw[fill=white] (.8,0) circle [radius=.1] node[black,font=\large,below=.1cm]{};
\draw[fill=black] (2.1,0) circle [radius=.1] node[black,font=\large,below=.1cm]{};
\end{tikzpicture}}
\caption{#2}
\label{#1}
\end{figure}}

\newcommand{\BSfig}[2]{\begin{figure}[h]
\centering
\subfloat{
\begin{tikzpicture}[scale=2, every node/.style={scale=1.3}]
\draw[very thick](-4,0)--(-2.1,0)node[black,left=1.2cm,above=.2cm]{$(0,1)$};
\draw[thick,dashed](-1.9,0)--(-.8,0)node[black,left=0.8cm,above=.3cm]{$H_1$};
\draw[very thick](-0.6,0)--(0.6,0)node[black,left=.8cm,above=.2cm]{$(0,1)$};
\draw[thick,dashed](0.8,0)--(1.9,0)node[black,left=0.8cm,above=.3cm]{$H_2$};
\draw[very thick](2.1,0)--(4,0)node[black,left=1.4cm,above=.2cm]{$(1,0)$};
\draw[fill=white] (-2,0) circle [radius=.1] node[black,font=\large,below=.1cm]{};
\draw[fill=white] (-0.7,0) circle [radius=.1] node[black,font=\large,below=.1cm]{};
\draw[fill=white] (0.7,0) circle [radius=.1] node[black,font=\large,below=.1cm]{};
\draw[fill=white] (2,0) circle [radius=.1] node[black,font=\large,below=.1cm]{};
\end{tikzpicture}}
\caption{#2}
\label{#1}
\end{figure}}

\newcommand{\BLfig}[2]{\begin{figure}[h]
\centering
\subfloat{
\begin{tikzpicture}[scale=2, every node/.style={scale=1.3}]
\draw[very thick](-4,0)--(-0.9,0)node[black,left=2.1cm,above=.2cm]{$(0,1)$};
\draw[thick,dashed](-.7,0)--(.7,0)node[black,left=1.0cm,above=.3cm]{$H$};
\draw[very thick](0.9,0)--(2.0,0)node[black,left=.8cm,above=.2cm]{$(n,1)$};
\draw[very thick](2.2,0)--(4,0)node[black,left=1.4cm,above=.2cm]{$(1,0)$};
\draw[fill=white] (-.8,0) circle [radius=.1] node[black,font=\large,below=.1cm]{};
\draw[fill=white] (.8,0) circle [radius=.1] node[black,font=\large,below=.1cm]{};
\draw[fill=black] (2.1,0) circle [radius=.1] node[black,font=\large,below=.1cm]{};
\end{tikzpicture}}
\caption{#2}
\label{#1}
\end{figure}}

\newcommand{\BHWBfig}[2]{\begin{figure}[h]
\centering
\subfloat{
\begin{tikzpicture}[scale=2, every node/.style={scale=1.3}]
\draw[very thick](-4,0)--(-2.1,0)node[black,left=1.2cm,above=.2cm]{$(0,1)$};
\draw[very thick](-1.9,0)--(-.8,0)node[black,left=0.8cm,above=.2cm]{$(1,0)$};
\draw[very thick](-0.6,0)--(0.6,0)node[black,left=.8cm,above=.2cm]{$(0,1)$};
\draw[thick, dashed](0.8,0)--(1.9,0)node[black,left=0.8cm,above=.3cm]{$H$};
\draw[very thick](2.1,0)--(4,0)node[black,left=1.4cm,above=.2cm]{$(1,0)$};
\draw[fill=black] (-2,0) circle [radius=.1] node[black,font=\large,below=.1cm]{};
\draw[fill=black] (-0.7,0) circle [radius=.1] node[black,font=\large,below=.1cm]{};
\draw[fill=white] (0.7,0) circle [radius=.1] node[black,font=\large,below=.1cm]{};
\draw[fill=white] (2,0) circle [radius=.1] node[black,font=\large,below=.1cm]{};
\end{tikzpicture}}
\caption{#2}
\label{#1}
\end{figure}}

\newcommand{\BHWDsfig}[2]{\begin{figure}[h]
\centering
\subfloat{
\begin{tikzpicture}[scale=2, every node/.style={scale=1.3}]
\draw[very thick](-4,0)--(-2.1,0)node[black,left=1.2cm,above=.2cm]{$(0,1)$};
\draw[very thick](-1.9,0)--(-.8,0)node[black,left=0.8cm,above=.2cm]{$(1,0)$};
\draw[thick, dashed](-0.6,0)--(0.6,0)node[black,left=.8cm,above=.3cm]{$H$};
\draw[very thick](0.8,0)--(1.9,0)node[black,left=0.8cm,above=.2cm]{$(0,1)$};
\draw[very thick](2.1,0)--(4,0)node[black,left=1.4cm,above=.2cm]{$(1,0)$};
\draw[fill=black] (-2,0) circle [radius=.1] node[black,font=\large,below=.1cm]{};
\draw[fill=white] (-0.7,0) circle [radius=.1] node[black,font=\large,below=.1cm]{};
\draw[fill=white] (0.7,0) circle [radius=.1] node[black,font=\large,below=.1cm]{};
\draw[fill=black] (2,0) circle [radius=.1] node[black,font=\large,below=.1cm]{};
\end{tikzpicture}}
\caption{#2}
\label{#1}
\end{figure}}

%% file: Bibliography.bib
@article{Chrusciel:2012jk,
    author = "Chrusciel, Piotr T. and Lopes Costa, Joao and Heusler, Markus",
    title = "{Stationary Black Holes: Uniqueness and Beyond}",
    eprint = "1205.6112",
    archivePrefix = "arXiv",
    primaryClass = "gr-qc",
    doi = "10.12942/lrr-2012-7",
    journal = "Living Rev. Rel.",
    volume = "15",
    pages = "7",
    year = "2012"
}

@article{Emparan:2001wk,
    author = "Emparan, Roberto and Reall, Harvey S.",
    title = "{Generalized Weyl solutions}",
    eprint = "hep-th/0110258",
    archivePrefix = "arXiv",
    reportNumber = "CERN-TH-2001-293",
    doi = "10.1103/PhysRevD.65.084025",
    journal = "Phys. Rev. D",
    volume = "65",
    pages = "084025",
    year = "2002"
}

@article{Harmark:2004rm,
    author = "Harmark, Troels",
    title = "{Stationary and axisymmetric solutions of higher-dimensional general relativity}",
    eprint = "hep-th/0408141",
    archivePrefix = "arXiv",
    doi = "10.1103/PhysRevD.70.124002",
    journal = "Phys. Rev. D",
    volume = "70",
    pages = "124002",
    year = "2004"
}

@article{Hollands:2007aj,
    author = "Hollands, Stefan and Yazadjiev, Stoytcho",
    title = "{Uniqueness theorem for 5-dimensional black holes with two axial Killing fields}",
    eprint = "0707.2775",
    archivePrefix = "arXiv",
    primaryClass = "gr-qc",
    doi = "10.1007/s00220-008-0516-3",
    journal = "Commun. Math. Phys.",
    volume = "283",
    pages = "749--768",
    year = "2008"
}

@article{Belinsky:1979mh,
    author = "Belinsky, V. A. and Sakharov, V. E.",
    title = "{Stationary Gravitational Solitons with Axial Symmetry}",
    journal = "Sov. Phys. JETP",
    volume = "50",
    pages = "1--9",
    year = "1979"
}

@article{Belinsky:1971nt,
    author = "Belinsky, V. A. and Zakharov, V. E.",
    title = "{Integration of the Einstein Equations by the Inverse Scattering Problem Technique and the Calculation of the Exact Soliton Solutions}",
    journal = "Sov. Phys. JETP",
    volume = "48",
    pages = "985--994",
    year = "1978"
}

@article{Chen:2008fa,
    author = "Chen, Yu and Teo, Edward",
    title = "{A Rotating black lens solution in five dimensions}",
    eprint = "0808.0587",
    archivePrefix = "arXiv",
    primaryClass = "gr-qc",
    doi = "10.1103/PhysRevD.78.064062",
    journal = "Phys. Rev. D",
    volume = "78",
    pages = "064062",
    year = "2008"
}

@article{Kunduri:2018qqt,
    author = "Kunduri, Hari K. and Lucietti, James",
    title = "{New thermodynamic identities for five-dimensional black holes}",
    eprint = "1810.13210",
    archivePrefix = "arXiv",
    primaryClass = "hep-th",
    doi = "10.1088/1361-6382/ab0982",
    journal = "Class. Quant. Grav.",
    volume = "36",
    number = "7",
    pages = "07LT02",
    year = "2019"
}

@article{Figueras:2009mc,
    author = "Figueras, Pau and Jamsin, Ella and Rocha, Jorge V. and Virmani, Amitabh",
    title = "{Integrability of Five Dimensional Minimal Supergravity and Charged Rotating Black Holes}",
    eprint = "0912.3199",
    archivePrefix = "arXiv",
    primaryClass = "hep-th",
    reportNumber = "DCPT-09-87, ULB-TH-09-43",
    doi = "10.1088/0264-9381/27/13/135011",
    journal = "Class. Quant. Grav.",
    volume = "27",
    pages = "135011",
    year = "2010"
}

@article{Kunduri:2014iga,
    author = "Kunduri, Hari K. and Lucietti, James",
    title = "{Black hole non-uniqueness via spacetime topology in five dimensions}",
    eprint = "1407.8002",
    archivePrefix = "arXiv",
    primaryClass = "hep-th",
    reportNumber = "EMPG-14-13",
    doi = "10.1007/JHEP10(2014)082",
    journal = "JHEP",
    volume = "10",
    pages = "082",
    year = "2014"
}

@article{Tomizawa:2016kjh,
    author = "Tomizawa, Shinya and Nozawa, Masato",
    title = "{Supersymmetric black lenses in five dimensions}",
    eprint = "1606.06643",
    archivePrefix = "arXiv",
    primaryClass = "hep-th",
    doi = "10.1103/PhysRevD.94.044037",
    journal = "Phys. Rev. D",
    volume = "94",
    number = "4",
    pages = "044037",
    year = "2016"
}

@article{Breunholder:2017ubu,
    author = {Breunh\"older, Veronika and Lucietti, James},
    title = "{Moduli space of supersymmetric solitons and black holes in five dimensions}",
    eprint = "1712.07092",
    archivePrefix = "arXiv",
    primaryClass = "hep-th",
    reportNumber = "EMPG-17-25",
    doi = "10.1007/s00220-018-3215-8",
    journal = "Commun. Math. Phys.",
    volume = "365",
    number = "2",
    pages = "471--513",
    year = "2019"
}

@article{Hollands:2008fm,
    author = "Hollands, Stefan and Yazadjiev, Stoytcho",
    title = "{A Uniqueness theorem for stationary Kaluza-Klein black holes}",
    eprint = "0812.3036",
    archivePrefix = "arXiv",
    primaryClass = "gr-qc",
    doi = "10.1007/s00220-010-1176-7",
    journal = "Commun. Math. Phys.",
    volume = "302",
    pages = "631--674",
    year = "2011"
}

@article{Lucietti:2020ltw,
    author = "Lucietti, James and Tomlinson, Fred",
    title = "{Moduli space of stationary vacuum black holes from integrability}",
    eprint = "2008.12761",
    archivePrefix = "arXiv",
    primaryClass = "gr-qc",
    reportNumber = "EMPG-20-16",
    month = "8",
    year = "2020"
}

@article{Alaee:2019qhj,
    author = "Alaee, Aghil and Khuri, Marcus and Kunduri, Hari",
    title = "{Existence and Uniqueness of Stationary Solutions in 5-Dimensional Minimal Supergravity}",
    eprint = "1904.12425",
    archivePrefix = "arXiv",
    primaryClass = "gr-qc",
    month = "4",
    year = "2019"
}

@article{Lucietti:2020phh,
    author = "Lucietti, James and Tomlinson, Fred",
    title = "{On the nonexistence of a vacuum black lens}",
    eprint = "2012.00381",
    archivePrefix = "arXiv",
    primaryClass = "gr-qc",
    reportNumber = "EMPG-20-23",
    doi = "10.1007/JHEP02(2021)005",
    journal = "JHEP",
    volume = "21",
    pages = "005",
    year = "2020"
}

@article{Kunduri:2017htl,
    author = "Kunduri, Hari K. and Lucietti, James",
    title = "{No static bubbling spacetimes in higher dimensional Einstein\textendash{}Maxwell theory}",
    eprint = "1712.02668",
    archivePrefix = "arXiv",
    primaryClass = "gr-qc",
    doi = "10.1088/1361-6382/aaa744",
    journal = "Class. Quant. Grav.",
    volume = "35",
    number = "5",
    pages = "054003",
    year = "2018"
}

@article{Gibbons:2002bh,
    author = "Gibbons, Gary W. and Ida, Daisuke and Shiromizu, Tetsuya",
    editor = "Maeda, K. and Sasaki, M.",
    title = "{Uniqueness and nonuniqueness of static vacuum black holes in higher dimensions}",
    eprint = "gr-qc/0203004",
    archivePrefix = "arXiv",
    reportNumber = "DAMTP-2002-28, RESCEU-1-02, UTAP-410",
    doi = "10.1143/PTPS.148.284",
    journal = "Prog. Theor. Phys. Suppl.",
    volume = "148",
    pages = "284--290",
    year = "2003"
}

@article{Hollands:2012cc,
    author = "Hollands, Stefan",
    title = "{Black hole uniqueness theorems and new thermodynamic identities in eleven dimensional supergravity}",
    eprint = "1204.3421",
    archivePrefix = "arXiv",
    primaryClass = "gr-qc",
    doi = "10.1088/0264-9381/29/20/205009",
    journal = "Class. Quant. Grav.",
    volume = "29",
    pages = "205009",
    year = "2012"
}

@article{Mizoguchi:1998wv,
    author = "Mizoguchi, Shunya and Ohta, Nobuyoshi",
    title = "{More on the similarity between D = 5 simple supergravity and M theory}",
    eprint = "hep-th/9807111",
    archivePrefix = "arXiv",
    reportNumber = "KEK-PREPRINT-98-99, OU-HET-297",
    doi = "10.1016/S0370-2693(98)01122-8",
    journal = "Phys. Lett. B",
    volume = "441",
    pages = "123--132",
    year = "1998"
}

@article{Maison:1979kx,
    author = "Maison, D.",
    title = "{Ehlers-Harrison type transformations for Jordan's extended theory of gravitation}",
    doi = "10.1007/BF00756907",
    journal = "Gen. Rel. Grav.",
    volume = "10",
    pages = "717--723",
    year = "1979"
}

@article{Pomeransky:2005sj,
    author = "Pomeransky, Andrei A.",
    title = "{Complete integrability of higher-dimensional Einstein equations with additional symmetry, and rotating black holes}",
    eprint = "hep-th/0507250",
    archivePrefix = "arXiv",
    doi = "10.1103/PhysRevD.73.044004",
    journal = "Phys. Rev. D",
    volume = "73",
    pages = "044004",
    year = "2006"
}

@article{Gibbons:2002ju,
    author = "Gibbons, Gary W. and Ida, Daisuke and Shiromizu, Tetsuya",
    title = "{Uniqueness of (dilatonic) charged black holes and black p-branes in higher dimensions}",
    eprint = "hep-th/0206136",
    archivePrefix = "arXiv",
    reportNumber = "DAMTP-2002-74, TIT-HEP-480",
    doi = "10.1103/PhysRevD.66.044010",
    journal = "Phys. Rev. D",
    volume = "66",
    pages = "044010",
    year = "2002"
}

@article{Tomizawa:2009ua,
    author = "Tomizawa, Shinya and Yasui, Yukinori and Ishibashi, Akihiro",
    title = "{Uniqueness theorem for charged rotating black holes in five-dimensional minimal supergravity}",
    eprint = "0901.4724",
    archivePrefix = "arXiv",
    primaryClass = "hep-th",
    doi = "10.1103/PhysRevD.79.124023",
    journal = "Phys. Rev. D",
    volume = "79",
    pages = "124023",
    year = "2009"
}

@article{Tomizawa:2009tb,
    author = "Tomizawa, Shinya and Yasui, Yukinori and Ishibashi, Akihiro",
    title = "{Uniqueness theorem for charged dipole rings in five-dimensional minimal supergravity}",
    eprint = "0911.4309",
    archivePrefix = "arXiv",
    primaryClass = "hep-th",
    doi = "10.1103/PhysRevD.81.084037",
    journal = "Phys. Rev. D",
    volume = "81",
    pages = "084037",
    year = "2010"
}

@article{Armas:2014gga,
    author = "Armas, Jay",
    title = "{Uniqueness of Black Holes with Bubbles in Minimal Supergravity}",
    eprint = "1408.4567",
    archivePrefix = "arXiv",
    primaryClass = "hep-th",
    doi = "10.1088/0264-9381/32/4/045001",
    journal = "Class. Quant. Grav.",
    volume = "32",
    number = "4",
    pages = "045001",
    year = "2015"
}

@article{Armas:2009dd,
    author = "Armas, Jay and Harmark, Troels",
    title = "{Uniqueness Theorem for Black Hole Space-Times with Multiple Disconnected Horizons}",
    eprint = "0911.4654",
    archivePrefix = "arXiv",
    primaryClass = "hep-th",
    doi = "10.1007/JHEP05(2010)093",
    journal = "JHEP",
    volume = "05",
    pages = "093",
    year = "2010"
}

@article{Ehlers:1959aug,
    author = "Ehlers, J.",
    editor = "Lichnerowicz, M. A. and Tonnelat, M. A.",
    title = "{Transformations of static exterior solutions of Einstein's gravitational field equations into different solutions by means of conformal mapping}",
    journal = "Colloq. Int. CNRS",
    volume = "91",
    pages = "275--284",
    year = "1962"
}

@article{Elvang:2004xi,
    author = "Elvang, Henriette and Emparan, Roberto and Figueras, Pau",
    title = "{Non-supersymmetric black rings as thermally excited supertubes}",
    eprint = "hep-th/0412130",
    archivePrefix = "arXiv",
    doi = "10.1088/1126-6708/2005/02/031",
    journal = "JHEP",
    volume = "02",
    pages = "031",
    year = "2005"
}

@inproceedings{Clement:2008qx,
    author = "Clement, Gerard",
    title = "{Sigma-model approaches to exact solutions in higher-dimensional gravity and supergravity}",
    booktitle = "{418th WE-Heraeus-Seminar}: {Models of Gravity in Higher Dimensions: From theory to Experimental search}",
    eprint = "0811.0691",
    archivePrefix = "arXiv",
    primaryClass = "hep-th",
    reportNumber = "LAPTH-1286-08",
    month = "11",
    year = "2008"
}

@article{Ida:2003wv,
    author = "Ida, Daisuke and Uchida, Yuki",
    title = "{Stationary Einstein-Maxwell fields in arbitrary dimensions}",
    eprint = "gr-qc/0307095",
    archivePrefix = "arXiv",
    doi = "10.1103/PhysRevD.68.104014",
    journal = "Phys. Rev. D",
    volume = "68",
    pages = "104014",
    year = "2003"
}

@article{Ford:2007th,
    author = "Ford, Jon and Giusto, Stefano and Peet, Amanda and Saxena, Ashish",
    editor = "Safarik, Karel and Sandor, Ladislav and Tomasik, Boris",
    title = "{Reduction without reduction: Adding KK-monopoles to five dimensional stationary axisymmetric solutions}",
    eprint = "0708.3823",
    archivePrefix = "arXiv",
    primaryClass = "hep-th",
    doi = "10.1088/0264-9381/25/7/075014",
    journal = "Class. Quant. Grav.",
    volume = "25",
    pages = "075014",
    year = "2008"
}

@article{Kunduri:2004da,
    author = "Kunduri, Hari K. and Lucietti, James",
    title = "{Electrically charged dilatonic black rings}",
    eprint = "hep-th/0412153",
    archivePrefix = "arXiv",
    reportNumber = "DAMTP-2004-149",
    doi = "10.1016/j.physletb.2005.01.044",
    journal = "Phys. Lett. B",
    volume = "609",
    pages = "143--149",
    year = "2005"
}

@article{Yazadjiev:2007cd,
    author = "Yazadjiev, Stoytcho S.",
    title = "{Black Saturn with dipole ring}",
    eprint = "0705.1840",
    archivePrefix = "arXiv",
    primaryClass = "hep-th",
    doi = "10.1103/PhysRevD.76.064011",
    journal = "Phys. Rev. D",
    volume = "76",
    pages = "064011",
    year = "2007"
}

@article{Lu:2008ze,
    author = "Lu, H. and Mei, Jianwei and Pope, C. N.",
    title = "{New Charged Black Holes in Five Dimensions}",
    eprint = "0806.2204",
    archivePrefix = "arXiv",
    primaryClass = "hep-th",
    reportNumber = "MIFP-08-12",
    doi = "10.1088/0264-9381/27/7/075013",
    journal = "Class. Quant. Grav.",
    volume = "27",
    pages = "075013",
    year = "2010"
}

@article{Chng:2008sr,
    author = "Chng, Brenda and Mann, Robert B. and Radu, Eugen and Stelea, Cristian",
    title = "{Charging Black Saturn?}",
    eprint = "0809.0154",
    archivePrefix = "arXiv",
    primaryClass = "hep-th",
    doi = "10.1088/1126-6708/2008/12/009",
    journal = "JHEP",
    volume = "12",
    pages = "009",
    year = "2008"
}

@article{Galtsov:2008zz,
    author = "Galtsov, Dmitri V.",
    editor = "Kenmoku, Masakatsu and Sasaki, Misao",
    title = "{Generating solutions via sigma-models}",
    eprint = "0901.0098",
    archivePrefix = "arXiv",
    primaryClass = "gr-qc",
    doi = "10.1143/PTPS.172.121",
    journal = "Prog. Theor. Phys. Suppl.",
    volume = "172",
    pages = "121--130",
    year = "2008"
}

@article{Kramer:1969wq,
    author = "Kramer, D. and Neugebauer, G.",
    title = "{An exact stationary solution of the einstein-maxwell equation. (in german)}",
    journal = "Annalen Phys.",
    volume = "24",
    pages = "59--61",
    year = "1969"
}

@article{Mizoguchi:1997si,
    author = "Mizoguchi, Shun'ya",
    title = "{E(10) symmetry in one-dimensional supergravity}",
    eprint = "hep-th/9703160",
    archivePrefix = "arXiv",
    reportNumber = "INS-1191",
    doi = "10.1016/S0550-3213(98)00322-8",
    journal = "Nucl. Phys. B",
    volume = "528",
    pages = "238--264",
    year = "1998"
}

@article{Julia:1980gr,
    author = "Julia, B.",
    title = "{Group Disintegrations}",
    reportNumber = "LPTENS-80-16",
    journal = "Conf. Proc. C",
    volume = "8006162",
    pages = "331--350",
    year = "1980"
}

@article{Iguchi:2007zz,
    author = "Iguchi, Hideo and Mishima, Takashi",
    editor = "Apostolopoulos, P. and Bona, Carles and Carot, J. and Mas, Ll. and Sintes, A. M. and Stela, J.",
    title = "{Solitonic generation of stationary and axisymmetric solutions of five-dimensional general relativity}",
    doi = "10.1088/1742-6596/66/1/012056",
    journal = "J. Phys. Conf. Ser.",
    volume = "66",
    pages = "012056",
    year = "2007"
}
